\documentstyle[psfig]{mn}
\title[High-redshift QSOs in FIRST]{High-redshift QSOs in the FIRST survey}
\author[C.R. Benn et al]{C.R. Benn,$^1$
\thanks{Email: crb@ing.iac.es}
M. Vigotti,$^2$ 
M. Pedani,$^3$
J. Holt,$^{1,4}$ 
K.-H. Mack,$^{6,2,7}$
R. Curran,$^{1,5}$
\newauthor
S.F. S\'{a}nchez$^1$ \\
$^1$Isaac Newton Group, Apartado 321, 38700 Santa Cruz de La Palma, Spain\\
$^2$Istituto di Radioastronomia, CNR, via Gobetti 101, 
40129 Bologna, Italy\\
$^3$Centro Galileo Galilei, 38700 Santa Cruz de La Palma, Spain \\
$^4$Department of Physics \& Astronomy, University of Sheffield, Hicks Building,
Sheffield S3 7RH, UK \\
$^5$Department of Physical Sciences,
University of Hertfordshire,
College Lane,
Hatfield,
Herts,
AL10 9AB, UK \\
$^6$ASTRON/NFRA, Postbus 2, NL-7990 AA Dwingeloo, the Netherlands \\
$^7$Radioastronomisches Institut der Universit\"{a}t Bonn,
Auf dem H\"{u}gel 71, D-53121 Bonn, Germany \\
}
\begin{document}
\maketitle

\begin{abstract}
In a pilot search for high-redshift radio QSOs, we have obtained 
spectra of 55
FIRST sources (S$_{1.4GHz} >$ 1 mJy) with 
very red (O-E $>$ 3) starlike optical identifications.
10 of the candidates are QSOs with redshifts 3.6 $< z <$ 4.4
(4 previously known), six with $z >$ 4.
The remaining 45 candidates comprise:
a $z =$ 2.6 BAL QSO;
3 low-redshift galaxies with narrow emission lines;
18 probable radio galaxies; and
23 M stars (mainly misidentifications).
The success rate (high-redshift QSOs / 
spectroscopically-observed candidates) for this search
is 1/2 for $S_{1.4GHz} >$ 10 mJy, 
and 1/9 for $S_{1.4GHz} >$ 1 mJy. 
With an effective search area of 4030 deg$^2$, the
surface density of high-redshift ($z>$ 4) QSOs
discovered with this
technique is
0.0015 deg$^{-2}$. 
\end{abstract}


\section{Introduction}
Observations of high-redshift ($z >$ 4) QSOs strongly
constrain models of:
the formation and evolution of galaxies
and their central black holes (Kauffmann \& Haehnelt 2000);
the contribution of QSOs to the
ionisation of the intergalactic medium at high redshifts
(Steidel, Pettini, Adelberger 2001);
and, through studies of
the Lyman absorption forest,
the chemical evolution of the intergalactic medium along the line
of sight to the QSO (Rauch 1998, Hamann \& Ferland 1999).

Most of the $\sim$ 300 $z >$ 4 QSOs now known have been discovered 
as a result of searches for objects
with unusually red optical colours (Fig. 1).
A summary of recent searches is given in Table 1.
Kennefick et al (1995a, b) searched for objects with unusually red
$g,r,i$ colours 
in 340 deg$^2$ covered by the second Palomar
Sky Survey (POSS-II) of the northern hemisphere
and discovered 10 $z>$ 4 QSOs (see also Djorgovski 2001).
Similarly, 
Sloan Digital Sky Survey (SDSS) commissioning images 
have been searched for objects with red $u,g,r,i,z$
colours.  More than 100 $z >$ 3.5 QSOs have been found so far
(Fan et al 2000, 2001; Schneider et al 2001;
Zheng et al  2001; Anderson et al 2001, Richards et al 2001), 
and this number will 
rise rapidly as SDSS 
nears completion
($\sim$ 10000 deg$^2$) 
over the next few years.
In the south, Storrie-Lombardi et al (2001) 
searched UKST plates scanned with the
Automated Plate Measuring facility (APM) in Cambridge
for objects with $B_J-R >$ 2.5, and found 
49 QSOs with $z >$ 4.

A disadvantage of the purely optical searches is that  
complex colour criteria are needed to exclude the much larger number 
of red stars, which makes it difficult to correct for incompleteness.
Alternatively, simple one-colour criteria can be used, but then a large
fraction of
the high-redshift QSOs is missed (compare in Table 1 the
surface densities attained for different types of optical selection).
Starting with a sample of radio QSOs
allows one to reduce the number of candidates
without resorting to complex colour criteria, and 
is less likely to bias the selection against dusty objects.
E.g. Hook et al (1998) sought red objects identified with
flat-spectrum QSOs with $S_{5GHz} >$ 25 mJy over
1600 deg$^2$, and found 6 with $z >$ 3 (none with $z>$ 4).
Snellen et al (2001) sought red objects identified with 
flat-spectrum radio sources $S_{5GHz} >$ 30 mJy 
in the Cosmic Lens All-Sky Survey (CLASS, Myers et al 2001)
over 6400 deg$^2$, and
found 4 QSOs with $z >$ 4, i.e. 1
$z >$ 4 QSO per 1600 deg$^2$.

Higher surface densities can be attained by  
observing the counterparts of fainter radio sources.
In this paper we report on a pilot search for high-redshift QSOs
amongst very red, $O-E >$ 3 (see Fig. 1), 
optical identifications of FIRST 
radio sources $S_{1.4GHz}>$ 1 mJy (with no selection on radio
spectral index).

There is no overlap between this search and the FIRST bright quasar survey
(White et al 2000, Becker et al 2001), 
whose selection criterion $O-E <$ 2
preclude detection of most high-redshift QSOs (none with $z >$ 3.7 were
catalogued), as do the colour selection criteria of most large
QSO surveys, e.g. the 2dF QSO survey (Croom et al 2001).

\begin{figure}
\centering
\psfig{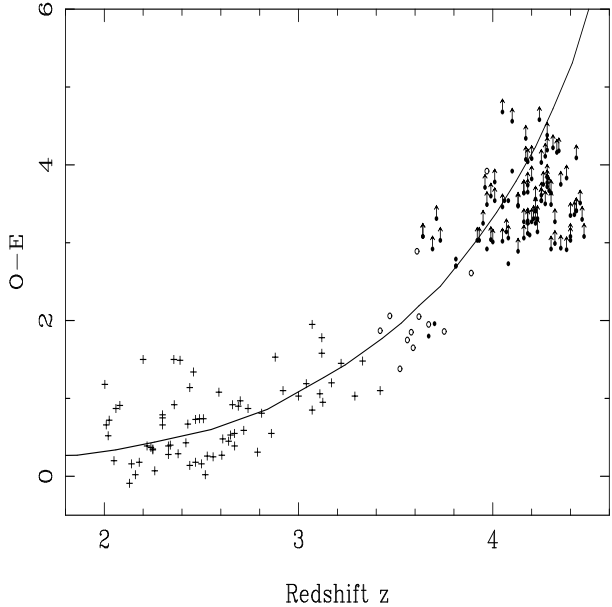}
\caption{The variation of POSSI $O-E$ colour with redshift, for 
QSOs from the FIRST bright quasar survey (crosses, White et al 2000),
the SDSS and POSSII searches for high-redshift QSOs
(spots), and a selection of radio QSOs with 3.4 $< z <$ 4.0 from
the NASA Extragalactic Database (circles).
The QSOs in these samples were selected in part on the basis of colour, but
nevertheless trace the typical reddening of a QSO with
increasing $z$ (prediction for typical 
QSO shown as solid line)
as the continuum redward
of the Ly$\alpha$ line moves out of the observed $O$ band (red limit 
$\approx$ 5000 \AA).  
As redshift increases above 4.5, the predicted brightness
of a QSO in $E$ band drops sharply, so few are likely to be detected.
With the criteria $O-E >$ 3, $E <$ 18.6 used here,
our QSO search is probably complete for $z >$ 4,
but incomplete for 3 $< z <$ 4.
}
\end{figure}

\begin{figure}
\centering
\psfig{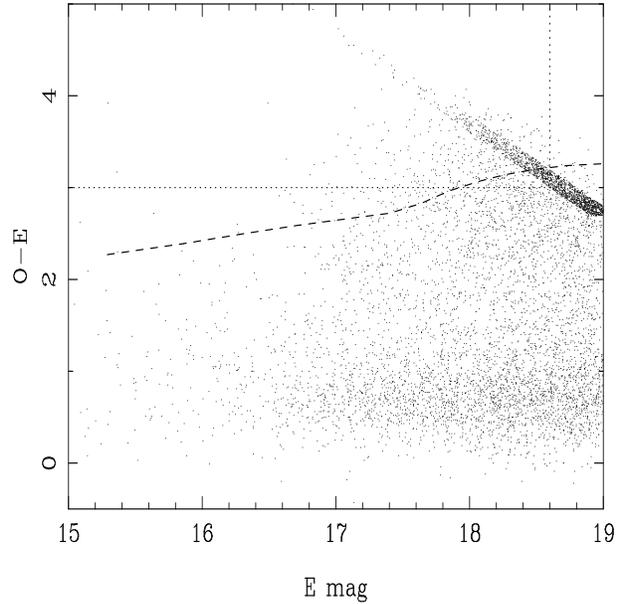}
\caption{Distribution in colour and magnitude of 
starlike (POSSI/APM) 
optical identifications of FIRST radio sources.
Objects with no $O$ magnitude have been plotted with
$O-E$ = 21.7 - $E$, and the $E$ magnitudes have been randomised
$\pm$ 0.1 so that the density of points in this region can be seen.
Most of the objects in the lower part of the diagram are true QSOs.
Most of those clustered at upper right are misclassified radio
galaxies (the dashed line shows the expected
colour-magnitude relationship for a redshifted giant elliptical
galaxy), 
and many of the other red objects are misidentifications
with galactic M stars.  Some may also be BAL QSOs, which tend to be red.
The candidate high-redshift QSOs observed here are drawn from the region
enclosed by the dotted line.
}
\end{figure}

\begin{figure}
\centering
\psfig{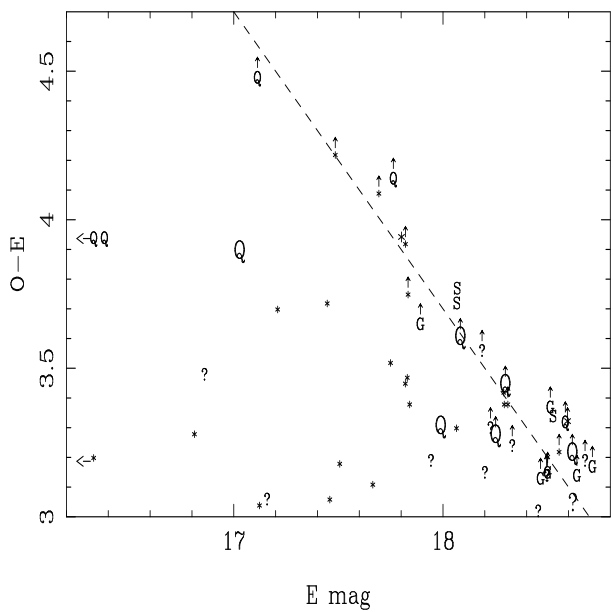}
\caption{Colour-magnitude distribution for the candidate
high-redshift QSOs.
The symbols for spectral type are as used in Table 2
(Q = QSO, S = narrow-emission-line galaxy, G = radio 
galaxy, 
$\ast$ = star, ? = probable radio galaxy).  
Large font indicates $S_{1.4GHz} >$ 10 mJy;
small font $S_{1.4GHz} <$ 10 mJy.
A nominal POSSI blue plate limit $O=$ 21.7 is indicated by the dashed line.
Objects
with no $O$ magnitude have been plotted with $O-E$ = 21.7 - $E$,
and the $E$ magnitudes have been randomised $\pm$ 0.2 mag to reduce
crowding.
The arrows indicate upper or lower limits.
}
\end{figure}

\begin{table*}
\begin{minipage}{170mm}
\caption{Recent searches for high-redshift QSOs}
\begin{tabular}{llllrrrrlll}
Search programme   &Colour    & optical & $S_{lim}$ & Area   &Cands&$z>$& N   &$\sigma$  & Hiz / & Reference\\
                   &criterion & limit   & mJy       & deg$^2$&     &    &     &deg$^{-2}$& cands &          \\
(1)                &(2)       &(3)      &  (4)      &(5)     &(6)  & (7)&(8)  &  (9)     & (10)   & (11) \\
DPOSS-II \rule{0mm}{3.5mm} 
                   &gri       &$r<$ 19.6& --        &  340   &  85 &4.0 &  10 & 0.029    & 0.11  &Kennefick et al 1995a,b\\
SDSS               &ugriz     &$i<$ 20  & --        &  180   &     &4.0 & 18  & 0.10     &       &Fan et al 2001 \\ 
APM/UKST           &$B_J-R>$2.5&$R<$ 19.3& --        & 8000   &     &4.0 & 49 & 0.0061    &        &Storrie-L. et al 2001\\   
INT wide-angle surv.
                   &griz      &$r<$ 24  & --        &   12\rlap{.5} 
                                                             &     &4.0 &  3 & 0.24      &        & Sharp et al 2001 \\     
CLASS flat-spectrum \rule{0mm}{3.5mm}         
                   &$O-E>$ 2  &$E<$19  &$S_5>$ 30   & 6400   & 27  &4.0 &  4 & 0.00063   &   0.15 & Snellen et al 2001\\
GB/FIRST flat-spec.        
                   &$O-E>$ 1.2&$E<$19.5&$S_{5}>$25  & 1100   & 50  &4.0 &  0 & 0.0        &   0.00 & Hook et al 1998   \\
``                 &$O-E>$ 1.2&$E<$19.5&$S_{5}>$25  & 1100   & 50  &3.0 &  6 & 0.005     &   0.12 & ``\\
FIRST              &$O-E>$ 3  &$E<$18.6&$S_{1.4}>$1 &4030    & 55  &4.0 &  6 & 0.0015    &   0.11 & This paper \\
``                 &$O-E>$ 3  &$E<$18.6&$S_{1.4}>$10&4030    &  9  &4.0 &  4 & 0.0010    &   0.44 & `` \\
\end{tabular}
Column 4 gives, where appropriate, the radio flux-density limit $S_\nu$, for frequency $\nu$
in GHz.
Column 9 gives the surface density of high-redshift QSOs discovered.
Column 10 gives the ratio between the number of high-redshift QSOs found (column 8) and
the number of candidates for spectroscopy (column 6).
\end{minipage}
\end{table*}

\section{Sample}
The FIRST radio catalogue (White et al 1997) currently includes 
722354 sources detected with peak
$S_{1.4GHZ} >$ 1 mJy over 7988 deg$^2$ ($\approx$ 90 sources deg$^{-2}$), 
mainly 7 $< RA <$ 17$^h$,
-5 $< Dec <$ 57$^o$.
We sought optical identifications of these sources with objects
catalogued by the APM 
(Irwin et al 1994) on the POSS-I O (blue) and E (red) plates.
We did not seek optical identifications at the mid-points of likely
double radio sources ($\sim$ 10\% of the catalogue), since this would
have required a substantial enlargement of the search area.
The colour-magnitude distribution of the star-like optical 
identifications lying within 1.5 arcsec of a FIRST source is shown
in Fig. 2.
We selected as candidate $z>$ 4 QSOs all starlike optical identifications
with $O-E >$ 3 (see Fig. 1) and 
$E <$ 18.6
(the POSS blue limit is $O \approx$ 21.7). 
Trial spectroscopy showed that $\sim$ 80\% of these are actually 
radio galaxies, which greatly outnumber QSOs at $E \sim$ 18,
$O-E \sim$ 3 (Fig. 2).
It was possible to filter out most of these by inspecting images
from the digitised POSS-II survey, using
the SExtractor program (Bertin \& Arnouts 1996) to
characterise the relationship between FWHM and intensity for stars
(in each image), and thus to distinguish between stellar and extended images.
In addition, we filtered out the $\approx$ 25\% of 
starlike images for which 
APM recorded no blue magnitude (i.e. implied magnitude fainter
than the blue plate limit), but which were easily visible 
on the Minnesota APS scans of POSS-I (Pennington et al
1993); these images were missed by APM due to the coarser 
scanning resolution
and greater susceptibility to confusion with nearby images.
This selection procedure yielded a total of 109 candidate high-redshift QSOs.

\section{Observations and reduction}
Spectra were obtained of a random sample of 55 of the 109 candidates
with the Isaac Newton Telescope's IDS spectrograph 
on 2001 Mar 14, and with the Isaac Newton (IDS),
William Herschel (ISIS) and Calar Alto
2.2-m (CAFOS) spectrographs on service nights 
between 2000 March 19 and 2001 Jan 5 (dates given in Table 2).
Apart from 2000 Apr 5, May 12 and 
2001 Jan 5, the nights were photometric.
The INT IDS spectrograph was used with the R150V grating, yielding
spectra with dispersion
6.5 \AA/pixel, usually centred at 6500 \AA.
The WHT ISIS dual-arm spectrograph was used with the R158R and R158B gratings
in the red and blue arms respectively (with a dichroic separating the
light blue and red of 6100 \AA).
In the red arm, with a TEK CCD detector, this yielded spectra with 
dispersion 2.9 \AA, usually centred at 7400 \AA.  
In the blue arm, with an EEV CCD, the dispersion
was 1.6 \AA/pixel, and the spectra were usually centred at 4800 \AA.
The CAFOS spectrum has dispersion 8.5 \AA/pixel.

The images were debiased and flat-fielded, cosmic rays were
eliminated, and the spectra were extracted, wavelength-calibrated
and intensity-calibrated in the usual way, using
the IRAF package.
A summary of the observations and results is given in Table 2.

\begin{figure*}
\centering
\psfig{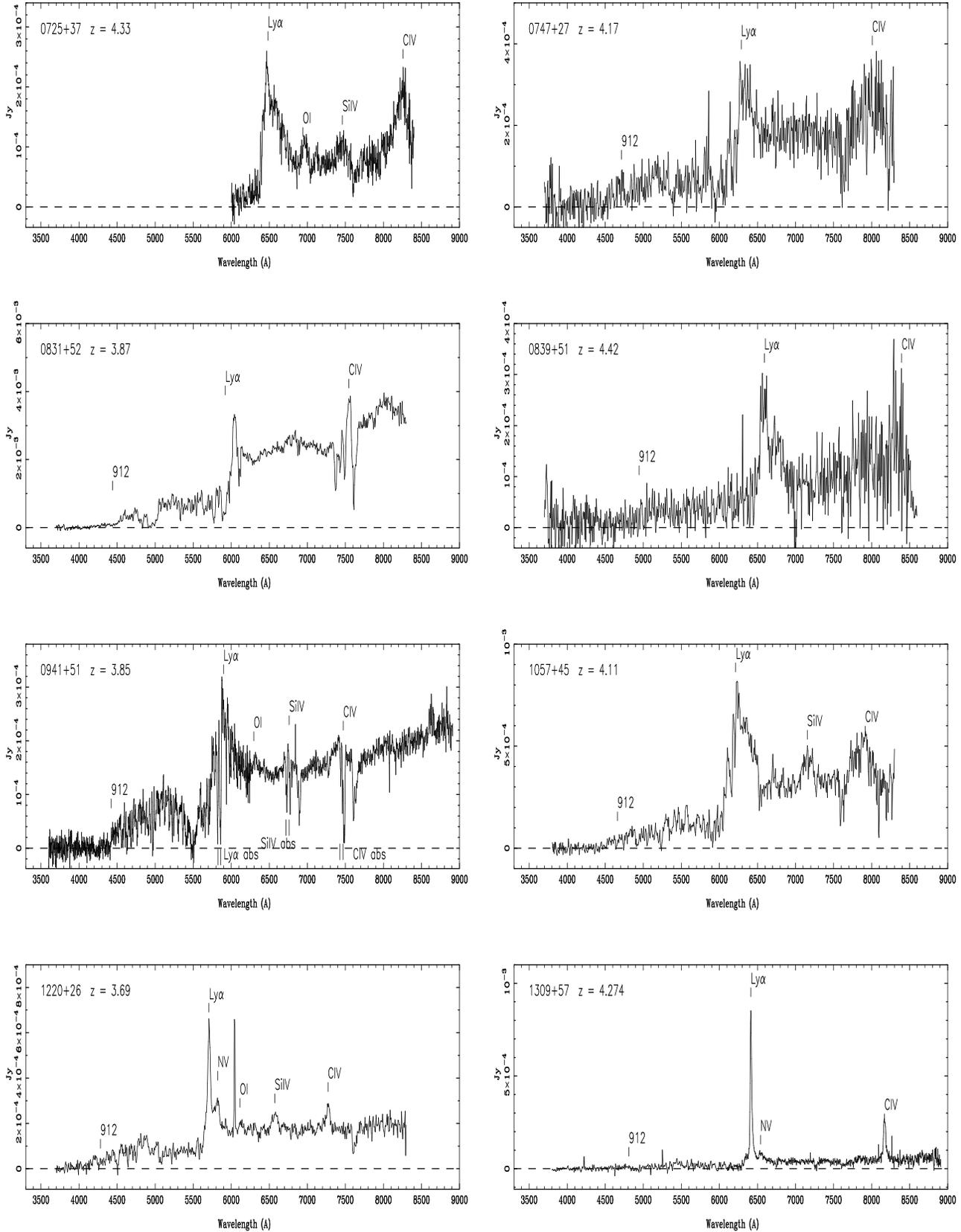}
\caption{
Spectra of 8 of the 10 high-redshift QSOs (see Fig 
5 for the remainder).
Spectral features are labelled at wavelengths corresponding to
the quoted redshift, assuming rest-frame wavelengths
in \AA\ of 1216 (Ly$\alpha$), 1240 (NV), 1302 
(OI/SiII blend), 1400 (SiIV/OIV] blend), 1549 (CIV).
For 0941+51, the metal-line absorption systems at $z$ = 3.80, 3.83 are also
indicated.
The spectra have not been corrected for the terrestrial atmospheric
absorption bands, notably at 7594 and 6867 \AA\ (A and B bands).
}
\end{figure*}

\begin{figure*}
\centering
\psfig{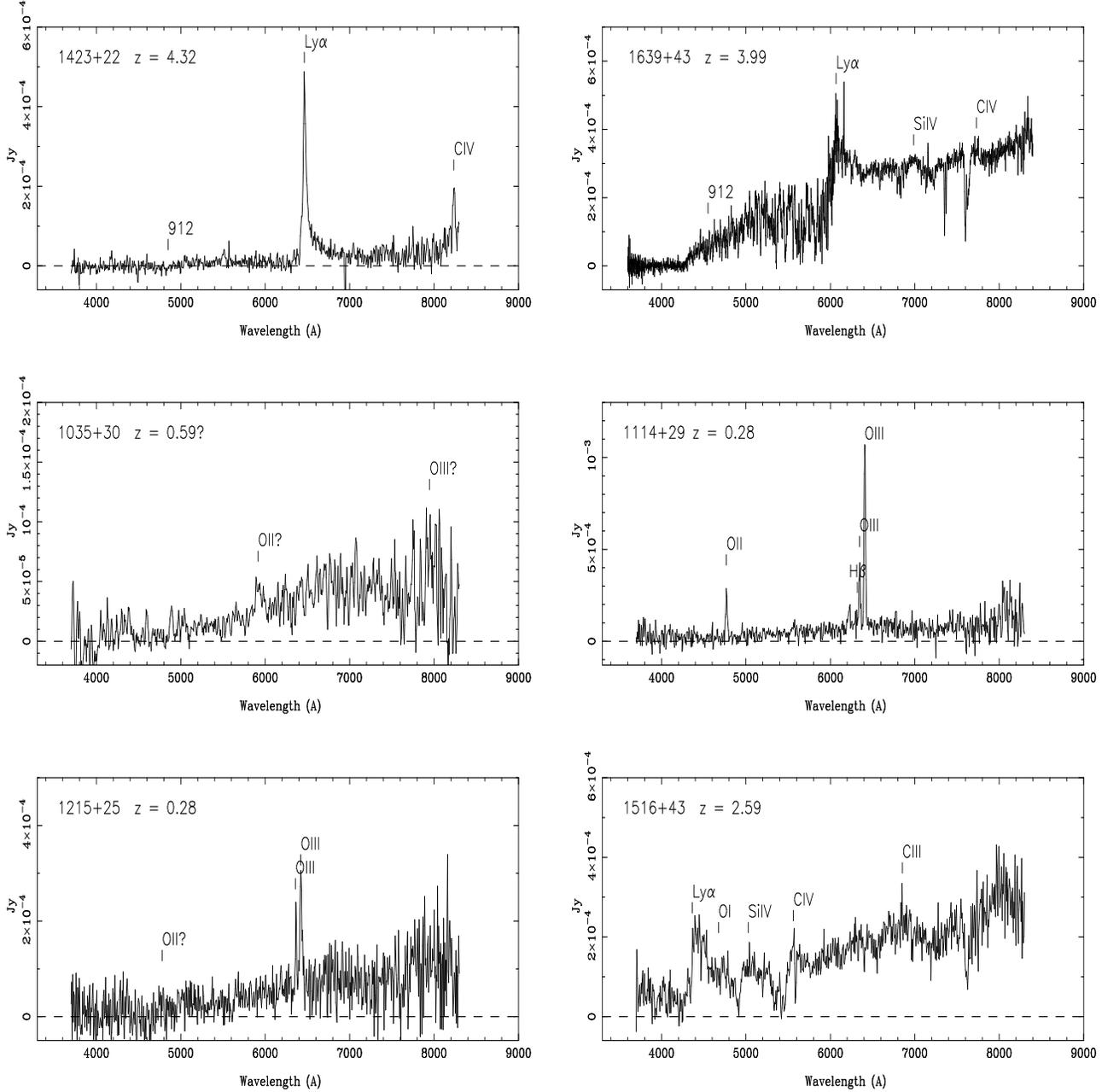}
\caption{
Spectra of the two remaining high-redshift QSOs (see caption of Fig. 4);
and of 4 other emission-line objects detected in this search (see text).
1516+43 was undetected on the POSS blue plate, but the field
is optically confused and its true colour may be bluer than $O-E$ = 3.
}
\end{figure*}

\begin{table*}
\begin{minipage}{170mm}
\caption{High-redshift QSO candidates}
\begin{tabular}{rrrrrrrrrrlll}
 RA   & Dec  & S$_{1.4}$  & R-O    & $E$ &$O-E$&Tel&Date& T &ID&z&$\sigma_z$  &Notes\\
J2000 & J2000&    mJy     & $\prime\prime$
                                   &     &     &   &obs.&sec&  &  & &     \\
(1) & (2) &(3)&(4)&(5)&(6)&(7)&(8)&(9)&(10)&(11)&(12)&(13) \\
\multicolumn{13}{l}{\bf High-redshift QSOs:} \\
\rule{0mm}{3.5mm}
072518.29&370517.9&  26.6&  0.6&  18.3&$>$3.4&W&010105&  900&Q & 4.33 &.01 &                              \\
074711.16&273903.6&   1.5&  0.8&  17.2&$>$4.5&I&010314&  200&Q & 4.17 &.02 &                              \\
083141.68&524517.1&   0.9&  0.7&  15.3&   3.9&I&010314&  200&Q & 3.87 &.03 &Irwin et al (1998), z = 3.91  \\
083946.11&511203.0&  41.6&  0.8&  18.5&$>$3.2&I&010314&  300&Q & 4.42 &.02 &Snellen et al (2000), z = 4.41\\
094119.36&511932.3&   2.7&  1.3&  17.6&$>$4.1&W&000319& 1200&Q & 3.85 &.01 &                              \\
105756.27&455553.0&   1.4&  0.5&  16.5&   3.9&I&000405&  300&Q & 4.11 &.02 &Stern et al (2000), z = 4.12  \\
122027.93&261903.6&  35.0&  0.4&  17.9&   3.3&I&010314& 1200&Q & 3.694&.005&                              \\
130940.61&573309.1&  11.5&  1.1&  18.1&$>$3.6&W&000319&  300&Q & 4.274&.005&                              \\
142308.19&224157.5&  35.4&  0.9&  18.5&$>$3.2&I&010314& 1200&Q & 4.316&.005&                              \\
151601.55&430931.3&   1.5&  0.9&  18.4&$>$3.3&I&010314&  300&Q & 2.590&.005&                              \\
163950.51&434003.3&  23.8&  0.5&  17.1&   3.9&W&000718& 1000&Q & 3.99 &.015&                              \\
\multicolumn{12}{l}{\bf Low-redshift QSOs and galaxies:} \\
\rule{0mm}{3.5mm}
091441.31&295621.4&   4.8&  0.3&  18.6&$>$3.1&I&010314&  300&G &      &    &                              \\
093255.83&353439.1&   1.9&  0.9&  18.2&   3.1&I&010314&  300&? &      &    &                              \\
101058.24&283247.0&   2.3&  0.4&  18.6&$>$3.1&I&010314&  100&? &      &    &                              \\
103549.91&300732.1&   9.0&  0.3&  18.4&   3.3&I&010314&  300&S & 0.587&.002&3727, 5007?                   \\
103647.80&420852.7&   2.2&  0.4&  18.6&$>$3.1&I&010314&  200&? &      &    &blue SED                      \\
104622.91&345436.0&   1.0&  0.7&  18.6&$>$3.1&I&010314&  300&G &      &    &                              \\
104801.61&260032.3&  11.2&  0.5&  18.6&   3.0&I&010314&  300&? &      &    &                              \\
104958.45&411043.4&   1.3&  0.5&  18.0&   3.2&I&010314&  300&? &      &    &                              \\
105719.80&342807.0&   5.4&  0.0&  18.1&$>$3.6&I&010314&  300&G &      &    &                              \\
105750.00&353300.4&   2.3&  0.7&  18.5&$>$3.2&I&010314& 1300&? &      &    &                              \\
111411.21&293244.4&   1.1&  1.0&  18.0&   3.8&I&010314&  300&S & 0.279&.001&3727 4861 4959 5007 6563      \\
111557.91&352757.1&   3.4&  1.4&  18.5&$>$3.2&I&010314&  300&? &      &    &                              \\
111707.05&411736.1&   0.7&  0.8&  18.5&$>$3.2&I&010314&  300&G &      &    &                              \\
112028.10&345829.4&   6.5&  0.8&  18.2&$>$3.5&I&010314&  300&? &      &    &possible high-redshift QSO    \\
115230.18&271808.4&   4.2&  0.4&  17.0&   3.0&I&010314&  300&? &      &    &                              \\
121532.16&250956.6&   1.0&  0.7&  18.0&   3.7&I&010314&  300&S & 0.282&.001&3727 4861 4959 5007           \\
124450.93&353906.7&   1.5&  0.2&  17.1&   3.5&I&000405&  300&? &      &    &                              \\
130604.97&352603.9&   1.1&  0.9&  18.4&   3.0&I&010314&  300&? &      &    &                              \\
134951.93&382334.9&   1.7&  0.4&  18.4&$>$3.3&I&010314&  300&? &      &    &possible high-redshift QSO    \\
140816.68&352205.8&   9.6&  1.1&  18.4&$>$3.4&I&010314&  300&G &      &    &                              \\
143749.26&325917.8&   1.3&  0.7&  18.6&$>$3.1&I&010314&  300&G &      &    &                              \\
\multicolumn{12}{l}{\bf M stars:} \\
\rule{0mm}{3.5mm}
072359.95&523949.4&   0.8&  0.8&  17.6&$>$4.1&I&010314&  300&* &      &    &                              \\
072801.49&380344.2&   1.6&  0.7&  17.3&   3.7&I&010314&  300&* &      &    &                              \\
074144.39&333549.3& 136.8&  0.6&  18.4&$>$3.3&W&010104&  400&* &      &    &                              \\
075127.64&272736.4&   1.2&  1.2&  18.0&$>$3.7&W&010104&  400&* &      &    &                              \\
084407.26&280740.4&   6.2&  0.5&  17.5&   3.0&W&010105& 1200&* &      &    &                              \\
085700.54&275540.0&   0.8&  0.8&  18.5&$>$3.2&W&010105&  400&* &      &    &                              \\
090556.33&224419.1&   2.4&  1.5&  17.5&$>$4.2&I&010314&  200&* &      &    &                              \\
091953.76&340906.7&   1.2&  0.8&  17.1&   3.0&I&010314&  200&* &      &    &                              \\
100125.11&370629.4&   1.1&  0.8&  16.8&   3.2&I&010314&  200&* &      &    &                              \\
105102.41&314914.5&   9.0&  0.4&  17.8&$>$3.9&I&010314&  300&* &      &    &                              \\
111212.69&431015.5&   1.4&  0.9&  17.2&   3.7&C&000512&  900&* &      &    &                              \\
112518.16&495158.8&   3.0&  1.1&  17.8&   3.4&I&010314&  300&* &      &    &                              \\
113043.07&514134.1&   4.2&  0.5&  17.7&   3.1&I&010314&  300&* &      &    &                              \\
120849.72&392907.1&   0.8&  1.2&  15.3&   3.2&I&010314&  300&* &      &    &                              \\
132903.23&323031.5&   5.8&  1.4&  18.1&   3.4&I&010314&  300&* &      &    &                              \\
133833.44&531642.1&   0.9&  0.7&  17.9&   3.3&I&010314&  300&* &      &    &                              \\
135357.42&343659.7&  98.0&  1.2&  17.8&   3.9&I&010314&  300&* &      &    &                              \\
150859.95&271431.0&   1.0&  0.4&  17.9&   3.4&I&010314&  300&* &      &    &                              \\
150938.97&434649.8&   1.9&  0.8&  18.4&$>$3.4&I&010314&  300&* &      &    &                              \\
161150.75&434412.4&   1.0&  1.0&  17.8&   3.4&I&010314&  300&* &      &    &                              \\
163535.34&352415.9&   5.2&  0.6&  18.1&   3.4&W&000718&  300&* &      &    &                              \\
164217.03&402230.8&   6.4&  0.2&  17.4&   3.1&W&000718&  300&* &      &    &                              \\
164401.60&420047.4&   4.2&  1.2&  17.6&   3.5&W&000718&  300&* &      &    &                              \\
\end{tabular}
The columns give: 
(1-2) right ascension and declination,
(3) FIRST 1.4-GHz (integrated, sometimes $<$ peak) flux density, 
(4) radio - optical displacement,
(5-6) APM POSSI $E$ magnitude and $O-E$ colour, 
(7) telescope with which spectrum
obtained (W = William Herschel 4.2-m, I = Isaac Newton 2.5-m,
C = Calar Alto 2.2-m), 
(8) date of observation,
(9) exposure time, 
(10) spectral type (Section 4),
(11-12) redshift and rms error on redshift.
For low-redshift emission-line objects, the rest-frame wavelengths of detected lines
are given in \AA\ in the last column: 3727 OII, 4861 H$\beta$,
4959/5007 OIII, 6563 H$\alpha$.
For high-redshift QSOs, see Figs. 4, 5.
\end{minipage}
\end{table*}

\section{Results}
Spectra showing clear emission or absorption features were classified
`Q' (QSO, broad emission lines),
`S' (Sy2 or starburst type spectrum, narrow emission lines) or
`$\ast$' (spectral features of M star).
The remainder have been classified 
`G' (probable radio galaxy, i.e. giant elliptical), where 
the characteristic 4000-\AA\ break is detected, otherwise
`?'.
Most of the `?' objects will be radio galaxies, rather than 
misidentifications with stars,
judging from
the number of M stars detected, the typical distribution
of spectral types at this magnitude,
and the magnitude distributions 
and the distributions on the sky of the `?' and `$\ast$' objects.
One of the `?' objects, 1036+42, has a flat SED inconsistent with it being
a radio galaxy; it may be a low-redshift QSO ($z <$ 2.2).
Of the other `?' objects, none has an SED consistent with it being a 
high-redshift QSO, except perhaps 1120+34 and 1349+38.
The redshifts of the high-redshift QSOs were measured from
emission lines other than Ly$\alpha$ (because of
the asymmetrical absorption), except in the case of 0839+51.
The colour-magnitude diagram of the candidates is shown in Fig. 3.
The spectra of the high-redshift QSOs are presented in Figs. 4, 5.

We find 10 high-redshift QSOs ($z >$ 3.6) out of 55 candidates.
These candidates were selected from 109 for the whole FIRST 
catalogue, covering 7988 deg$^2$, so the effective area covered by this
search is $\sim$ 4030 deg$^2$, i.e. we find 
0.0025 high-redshift QSOs
deg$^{-2}$ with $z >$ 3.6,
or 0.0015 
deg$^{-2}$ with $z>$ 4.0.  
For $S_{1.4GHz}>$ 1 mJy, $E<$ 18.6, the $z>$ 4 search will be
near-complete (Fig. 1).
The surface density on the sky is compared with that from other
searches for high-redshift QSOs in Table 1.
The surface density of $z >$ 4 radio QSOs discovered here is 
twice as high as that from the Snellen et al search, because of the
lower flux-density limit.
The efficiency of the search
(number of high-redshift QSOs / number of candidates for spectroscopy) 
is 0.44 for $S_{1.4GHz}>$ 10 mJy, higher than for previous searches,
probably because of the careful filtering of the candidates.

Four of the 10 high-redshift QSOs were previously known:
0747+27 (M. Irwin, private communication),
0831+52 (Irwin et al 1998), 
0839+51 (Snellen et al 2001) and
1057+45 (Stern et al 2000).
0831+52 is unusually bright ($E$ = 15) and the 
Ly$\alpha$ forest been studied in detail
(Ellison et al 1999).
327-MHz flux densities are available for three of the high-redshift QSOs 
from the WENSS survey (Rengelink et al 1997).  The 0.3 - 1.4-GHz spectral
indices $\alpha$ ($S_\nu \propto \nu^\alpha$) are all flat:
-0.1 for 0725+37,
 0.0 for 0839+51, and 
-0.4 for 1639+43.
The other 7 QSOs are too faint to be detected in WENSS, or lie below
the WENSS declination limit.

The SEDs of the high-redshift radio QSOs reported here
are as varied as those found in optical searches.
At least 0941+51 and 0831+52 show metal-line absorption 
redward of Ly$\alpha$, probably associated with saturated Ly$\alpha$
lines in the forest.
One, 1309+57, shows very narrow Ly$\alpha$ emission, FWHM 20 \AA.
1639+43 has almost no Ly$\alpha$ emission, which is unusual (only $\sim$ 
1/60 of SDSS QSOs),
and it is similar in this respect to the $z =$ 4.2 radio QSO, 0918+06,
discovered
by Snellen et al.  Snellen et al suggested that the unusually strong 
absorption of the Ly$\alpha$ line might be due to the host galaxy
(in both cases this is probably a giant elliptical).
A few of the QSOs have probable damped Ly$\alpha$ systems (DLAs,
seen in $\sim$ 20\% of high-redshift QSOs).
0941+51 has a DLA at $z =$ 3.52, and this has now been observed at 
high spectral resolution with WHT ISIS
(Centurion et al, in preparation).

The other 45 candidates are a mixed bag (see Magliocchetti et al 2000
for the typical distribution of identification types of all colours
at 1 mJy).
1516+43 is a $z=$ 2.59 broad-absorption-line (BAL) QSO with
broad, deep SiIV and CIV absorption troughs blueward of the emission
features (Fig. 5).
1035+30 (probably), 1114+29 and 1215+25 are low-redshift galaxies
exhibiting narrow emission lines (Fig. 5).
18 have SEDs consistent with those of 
giant elliptical galaxies at $z\sim$ 0.3.
23 have spectra of late-type M stars.
This number is consistent with that expected
for chance coincidences, but a few may be true
radio stars (Helfand et al 1999 describe a search for counterparts
of bright stars in FIRST).

\section{Conclusions}
We have conducted a pilot search 
for high-redshift radio QSOs, obtaining 
optical spectra of 55
FIRST radio sources (S$_{1.4GHz} >$ 1 mJy) with 
red (O-E $>$ 3) starlike optical identifications.
10 of the candidates are QSOs with redshifts 3.6 $< z <$ 4.4
(4 previously known).  Six have $z >$ 4.
The remaining 45 candidates are a mixture of low-redshift galaxies
(misclassified as stellar) and M stars (mainly misidentifications).
The success rate (high-redshift QSOs / candidates) for this search
is 1/2 for $S_{1.4GHz} >$ 10 mJy, 
and 1/9 for $S_{1.4GHz} >$ 1 mJy. 
With an effective search area of 4030 deg$^2$, the
surface density of $z>$ 4 QSOs discovered with this
technique is
0.0015 deg$^{-2}$. 

\vspace{3mm}
{\bf Acknowledgments}\\
We are grateful to Alessandro Caccianiga (Observatorio
de Lisboa) for taking spectra with the Calar
Alto 2.2-m telescope.
Rachel Curran and Joanna Holt were 1-year placement students at
ING in 1999-2000 and 2000-2001 respectively.
KHM was supported by the European Commission, TMR Programme, Research
Network Contract ERBFMRXCT96-0034 ``CERES''.
The William Herschel and Isaac Newton Telescopes are operated on the
island of La Palma by the Royal Greenwich Observatory in the Spanish 
Observatorio del Roque de los Muchachos of the
Instituto de Astrofisica de Canarias.
The Calar Alto observatory is 
operated by the Max-Planck-Institute for Astronomy,
Heidelberg, jointly with the Spanish National Commission for Astronomy.
The APS Catalog of POSS I (http://aps.umn.edu) is supported by
NASA and the University of Minnesota.
The NASA/IPAC Extragalactic Database (NED)
is operated by the Jet Propulsion Laboratory, California
Institute of Technology, under contract with NASA.

\end{document}